\documentclass[aps,apl,reprint,superscriptaddress
]{revtex4-2}
\usepackage{graphicx}
\usepackage{amsmath,amsfonts,amssymb}	
\usepackage{mathrsfs}
\usepackage[normalem]{ulem}
\usepackage{xcolor}
\usepackage{epstopdf}
\usepackage[retain-explicit-plus]{siunitx}
\usepackage[version=4]{mhchem}
\usepackage{graphicx}
\usepackage{subfigure}
\usepackage{braket}
\usepackage{booktabs}
\usepackage{hyperref}
\hypersetup{
    colorlinks=true,
    linkcolor=blue,
    filecolor=magenta,      
    urlcolor=blue,
    citecolor=blue
    }
	\definecolor{alizarin}{rgb}{0.82, 0.1, 0.26} 

\draft 

\begin{document}


    \title
  {A figure-of-merit-based framework to evaluate photovoltaic materials}



\author{Andrea Crovetto}
\email[]{E-mail: ancro@dtu.dk}
\affiliation{National Centre for Nano Fabrication and Characterization (DTU Nanolab), Technical University of Denmark, 2800 Kongens Lyngby, Denmark}

\begin{abstract}
I propose a general quantitative framework to evaluate the quality, track the historical development, and guide future optimization of photovoltaic (PV) absorbers at any development level, both lab-made and computer-simulated. The framework is centered around a PV figure of merit designed to include efficiency limitations that are not captured by classic detailed balance methods derived from the Shockley-Queisser limit. A more stringent set of figure-of-merit-driven efficiency limits are calculated for 28 experimentally synthesized PV absorbers and 10 PV computationally modeled absorbers. Among early-stage absorbers, this analysis reveals very large differences in their likelihood of achieving high PV efficiencies in the future. Since the proposed figure of merit is instantly evaluated from a single equation, it can be a suitable objective function for closed-loop research on PV materials in autonomous labs, while also providing a quantitative bridge between computationally determined material properties and PV efficiency.
\end{abstract}


\maketitle 



\section{Introduction}
The pace of exploration of new photovoltaic (PV) absorber materials has accelerated in recent years, owing in part to the popularization of high-throughput and autonomous materials research methods.~\cite{zhangSelfDrivenAutonomousMaterial2024,shengHighthroughputMicroscaleBandgap2025,mittmannPhosphosulfideSemiconductorsOptoelectronics2024,almoraDevicePerformanceEmerging2026,vermangEmergingInorganicSolar2026} The holy grail of PV materials research is a new photoabsorber material with similar performance and process tolerance to halide perovskites but without their stability and toxicity drawbacks.
Progress towards this goal is hampered by several fundamental and practical issues. How do we quantify the efficiency potential of new absorbers early in their development stage? Which material properties of an emerging absorber require optimization most urgently? Should we prioritize work on the absorber itself, or on the device surrounding it? Which emerging absorbers are worth investigating further, and which ones might we want to drop altogether?

These problems are fundamentally linked to the methods available to estimate the maximum PV efficiency of semiconducting absorbers. The physical principle of detailed balance between photon absorption and emission is the cornerstone of the most influential efficiency limits, which consist of the classic Shockley-Queisser (SQ) limit~\cite{Shockley1961,Guillemoles2019} and subsequent extensions to include the effect of non-radiative recombination~\cite{Yu2012,Blank2017,Kirchartz2018,kimInitioCalculationDetailed2021}.
Even though these methods are powerful predictors of open-circuit voltage limits in solar cells, they do not consider efficiency limitations arising from imperfect carrier collection, which often occurs under various combinations of carrier mobilities, doping density, and dielectric constant~\cite{Mattheis2007,kirchartzDetailedBalanceTheory2008}. Carrier collection losses affect short circuit current densities and fill factors but are difficult to model in a unified manner~\cite{Blank2017,Mattheis2007,kirchartzDetailedBalanceTheory2008}. Thus, detailed-balance methods provide upper bounds to the maximum efficiency achievable by a real PV absorber.

\begin{figure*}[t!]
\centering%
\includegraphics[width=\textwidth]{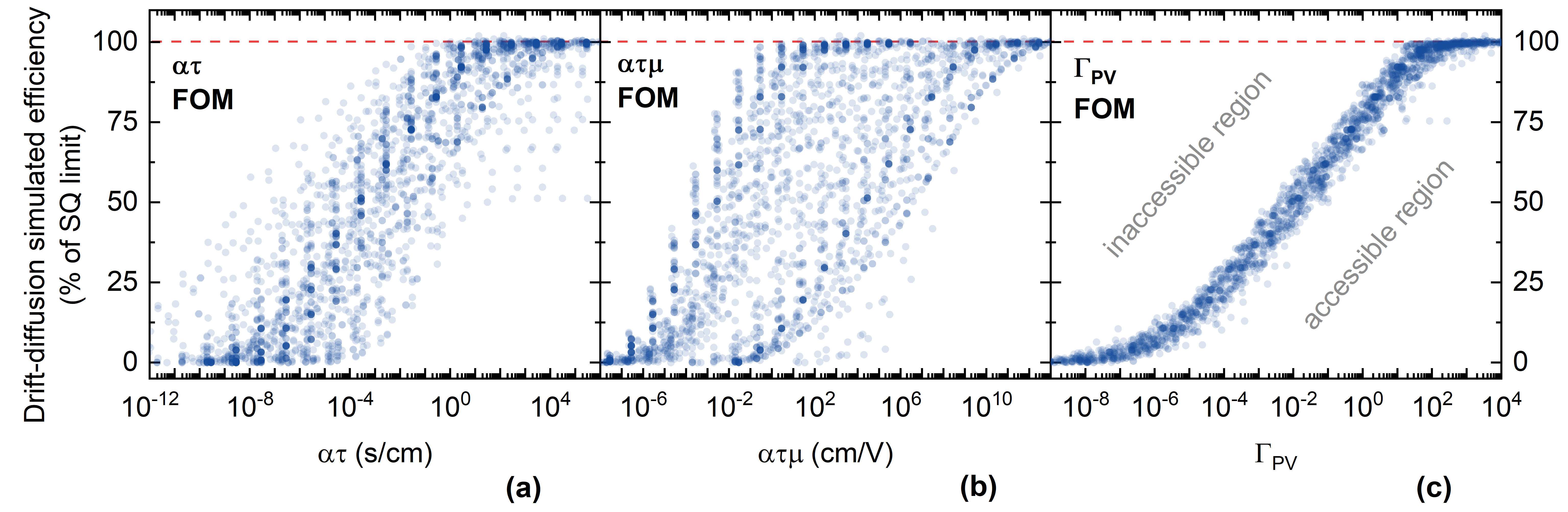}
\caption{Predictive ability of the $\alpha \tau$ (a), $\alpha \tau \mu$ (b), and $\Gamma_\mathrm{PV}$ figures of merit (c). The maximum efficiency of over 2500 hypothetical absorbers with different combinations of the $\alpha$, $\sigma$, $\tau$, $\mu$, $n$, $\epsilon$, $m$, and $E_\mathrm{g}$ properties (as obtained by drift-diffusion simulation) is plotted against the three FOMs. These data points constituted a training set in the original development of the $\Gamma_\mathrm{PV}$ FOM~\cite{crovettoPhenomenologicalFigureMerit2024}. Efficiencies are expressed as fraction of the SQ limit, indicated by a red dashed line. The $\Gamma_\mathrm{PV}$ FOM splits the efficiency versus FOM space into an accessible and an inaccessible region.}
\label{fig:different_foms}
\end{figure*}

To address this problem, here I propose an alternative route to the definition of efficiency limits. Rather than deducing them from fundamental physics (as in detailed balance methods), I suggest an inductive methodology based on the recently developed $\Gamma_\mathrm{PV}$ photovoltaic figure of merit (FOM)~\cite{crovettoPhenomenologicalFigureMerit2024}. The $\Gamma_\mathrm{PV}$ FOM provides a general quantitative framework to evaluate the quality, track the historical development, and guide future optimization of a generic PV absorber at any development level.

\section{Methods}

The $\Gamma_\mathrm{PV}$ FOM is a unitless function of eight properties of the PV absorber's bulk:

\begin{equation} \label{eq:fom_parameters}
\Gamma_\mathrm{PV} = f(\alpha,\sigma,\tau,\mu,n,\epsilon,m,E_\mathrm{g})
\end{equation}

Here, $\alpha$ and $\sigma$ are parameters describing the spectral average and spectral dispersion of the absorption coefficient spectrum, $\tau$ is the non-radiative recombination lifetime, $\mu$ is the carrier mobility, $n$ is the carrier concentration, $\epsilon$ is the static dielectric constant, $m$ is the density-of-states (DOS) effective mass, and $E_\mathrm{g}$ is the band gap. Definitions of these properties are given in Sec.~S3 and Ref.~\cite{crovettoPhenomenologicalFigureMerit2024}, including potential strategies to deal with direction- and carrier-type dependence.

The $\Gamma_\mathrm{PV}$ FOM is defined as:

\begin{equation} \label{eq:fom_definition}
\Gamma_\mathrm{PV} = {E_\mathrm{g}}^{2.5}\left(\cfrac{\mathcal{A}_1 \mathcal{A}_2 \mathcal{D}_1}{\mathcal{D}_2 \mathcal{D}_3 \mathcal{D}_4 \, (1+\mathcal{T}_1 \mathcal{T}_2 \mathcal{T}_3) \, (1+\mathcal{S}_1 \mathcal{S}_2)}\right)^{{E_\mathrm{g}}^{-0.8}}
\end{equation}

The various factors $\mathcal{A}_1, \mathcal{A}_2, \mathcal{D}_1 ... \, \mathcal{S}_2$ in Eq.~\ref{eq:fom_definition} are expressions~\cite{crovettoPhenomenologicalFigureMerit2024} containing the eight properties listed in Eq.~\ref{eq:fom_parameters}.

The value of $\Gamma_\mathrm{PV}$ can be calculated for any candidate PV absorber at any development level, as long as the eight bulk properties in Eq.~\ref{eq:fom_parameters} have been measured, calculated, or estimated. Once $\Gamma_\mathrm{PV}$ has been determined for a certain absorber, the maximum PV efficiency achievable by that absorber in a planar single-junction solar cell under the the standard 1-Sun, AM1.5G spectrum can be estimated as $\eta_\mathrm{\Gamma}$, defined as:

\begin{equation} \label{eq:eta_from_fom}
\eta_\mathrm{\Gamma}(\Gamma_\mathrm{PV},\eta_\mathrm{sq}) = \cfrac{\eta_\mathrm{sq}}{\left(1+\cfrac{k_1 \, {\Gamma_\mathrm{PV}}^{-0.235}}{1+k_2 \, {\Gamma_\mathrm{PV}}^{0.869}}\right)\left(1+k_3 \, {\Gamma_\mathrm{PV}}^{-0.362}\right)  }
\end{equation}

Here, $\eta_\mathrm{sq}$ is the Shockley-Queisser limiting efficiency of the absorber, which is readily calculated from $E_\mathrm{g}$~\cite{Shockley1961}, and $k_1, k_2, k_3$ are fixed parameters~\cite{crovettoPhenomenologicalFigureMerit2024}. For easier comparison between different PV absorbers with different band gaps, it is often convenient to express $\eta_\mathrm{\Gamma}$ as a fraction of the Shockley-Queisser limit. This SQ-normalized efficiency is defined as $\eta_\mathrm{\Gamma,sq} = \eta_\mathrm{\Gamma}/\eta_\mathrm{sq}$. 

\section{Results and discussion}

\begin{figure*}[t!]
\centering%
\includegraphics[width=0.9\textwidth]{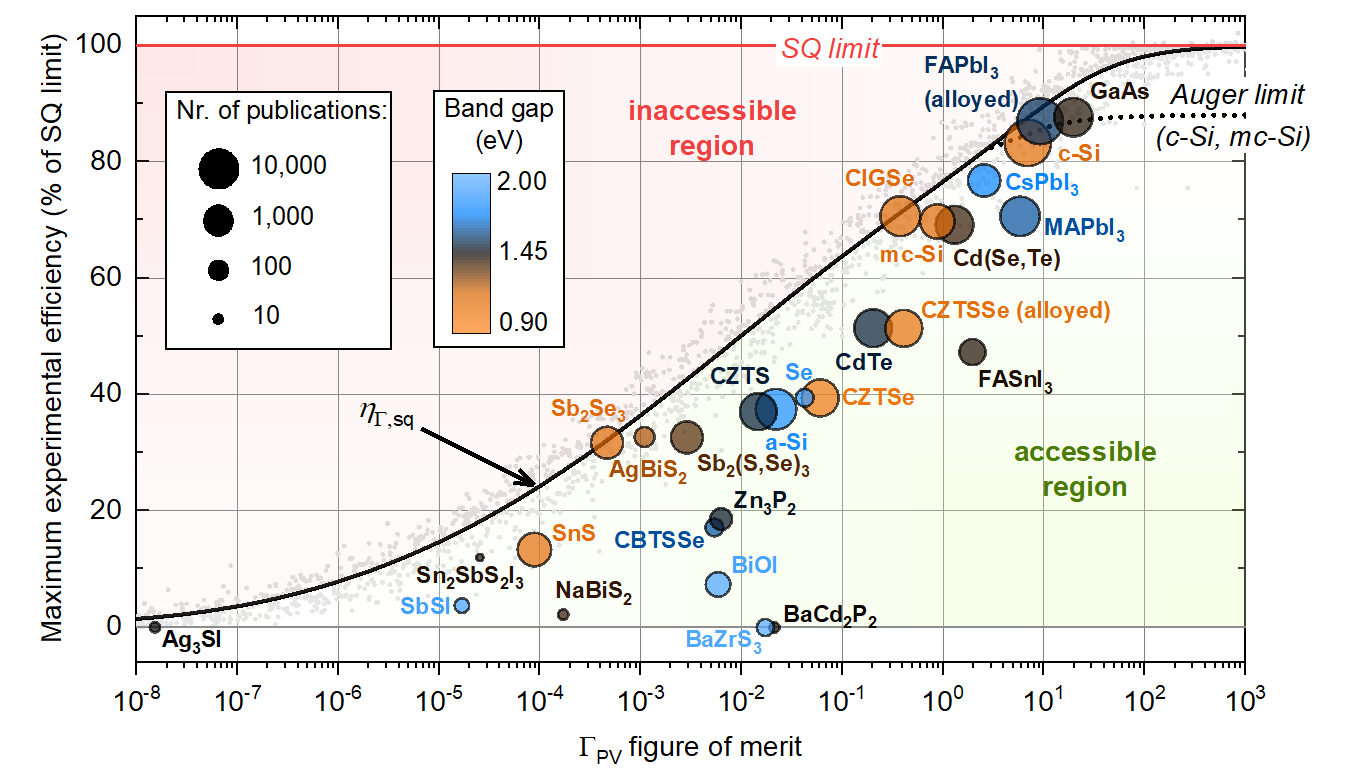}
\caption{Performance status of 28 experimentally synthesized PV absorbers, expressed as record efficiency (y-axis, normalized to SQ limit) versus $\Gamma_\mathrm{PV}$ FOM (x-axis). Record solar cells for all materials except SbSI have a planar architecture. The thick black line is the $\eta_\mathrm{\Gamma,sq}$ efficiency limit (Eq.~\ref{eq:eta_from_fom}). The dotted line indicates the Auger limit for c-Si and mc-Si (more stringent than their SQ limit). In this representation, the SQ limit is always 100\%. The size of the data points is proportional to the logarithm of the number of publications referring to each absorber in a PV context. The color of the data points indicates the band gap of each absorber. The small light grey data points are the training set used to derive the $\Gamma_\mathrm{PV}$ FOM (identical to the data points in Fig.~\ref{fig:different_foms}(c)). They give a visual impression of the $\pm 3.4\%$ absolute error bar of $\eta_\mathrm{\Gamma,sq}$ as an estimator of efficiency limits. The material properties used to calculate $\Gamma_\mathrm{PV}$ for each absorber are listed in Table~S2, Supporting Information. The maximum experimental efficiencies used for each material are listed in Table~S4, Supporting Information.}
\label{fig:fom_real_materials}
\end{figure*}

The predictive ability of $\Gamma_\mathrm{PV}$ is shown in Fig.~\ref{fig:different_foms}, compared to the case of the simpler $\alpha \tau$ and $\alpha \tau \mu$ FOMs~\cite{kaienburgHowSolarCell2020}.
The three plots in Fig.~\ref{fig:different_foms} show the maximum SQ-normalized efficiency of over 2500 hypothetical absorbers (as obtained by explicit drift-diffusion simulation) versus the three FOMs calculated for the same absorbers.
In the simulation~\cite{crovettoPhenomenologicalFigureMerit2024} the only PV losses besides radiative recombination are due to non-optimal values of the eight bulk properties used to define $\Gamma_\mathrm{PV}$ (Eq.~\ref{eq:fom_parameters}). It is apparent that the $\alpha \tau$ and $\alpha \tau \mu$ FOMs are generally not predictive of efficiency, although they can be useful indicators under specific conditions~\cite{kaienburgHowSolarCell2020,crovettoPhenomenologicalFigureMerit2024}. On the other hand, the $\Gamma_\mathrm{PV}$ FOM has higher predictive power because the simulated data roughly follows a single function that approaches zero for low $\Gamma_\mathrm{PV}$ values and $\eta_\mathrm{sq}$ for high $\Gamma_\mathrm{PV}$ values (Fig.~\ref{fig:different_foms}(c)). This function is just the SQ-normalized version of the $\eta_\mathrm{\Gamma}$ function defined in Eq.~\ref{eq:eta_from_fom} and it is shown in Fig.~\ref{fig:fom_real_materials} as a black line.

The $\Gamma_\mathrm{PV}$ FOM naturally leads to the definition of an "accessible region" (below the $\eta_\mathrm{\Gamma,sq}$ curve in Fig.~\ref{fig:fom_real_materials}) which is the efficiency range that PV absorbers can realistically access as a function of their figure of merit. The "inaccessible region" above the $\eta_\mathrm{\Gamma,sq}$ curve would be allowed by the SQ limit alone, but is not allowed by the additional losses introduced by non-optimal values of the eight bulk properties. The error bar of the $\eta_\mathrm{\Gamma,sq}$ efficiency limit is $\pm 3.4\%$ absolute~\cite{crovettoPhenomenologicalFigureMerit2024}, under the assumption that the values of $\alpha$, $\sigma$, $\tau$, $\mu$, $n$, $\epsilon$, $m$, and $E_\mathrm{g}$ used to calculate $\Gamma_\mathrm{PV}$ are known exactly. This corresponds to an absolute error around $\pm 1.1\%$ on the actual $\eta_\mathrm{\Gamma}$ efficiency (not SQ-normalized) for PV absorbers in the \SIrange{1.1}{1.5}{eV} band gap range, and less for absorbers outside this range.

In Fig.~\ref{fig:fom_real_materials}, the maximum (SQ-normalized) experimental efficiency $\eta_\mathrm{exp,sq}$ demonstrated by 28 emerging and established PV absorbers in planar solar cell configurations is plotted as a function of their corresponding $\Gamma_\mathrm{PV}$ FOM. Criteria and priorities applied to choose values of material properties from the literature are given in Sec.~S3 of the SI. Remarks on the specific values of the eight properties chosen to calculate $\Gamma_\mathrm{PV}$ for each of these materials are available in Sec.~S6 of the SI. 

As expected, all absorbers lie in the accessible region defined by the $\Gamma_\mathrm{PV}$ FOM. Several absorbers lie within error bar of the $\eta_\mathrm{\Gamma,sq}$ limit, indicating that solar cell architectures, contact layers, and interfaces are sufficiently optimized and do not introduce significant losses at the current level of bulk material quality in these absorbers.
GaAs is the absorber with both the highest FOM (20) and the maximum SQ-normalized experimental efficiency (88\%). Alloyed perovskites based on CH$_5$N$_2$PbI$_3$ (triple-cation and double-anion) have the second highest FOM, followed by c-Si and CH$_3$NH$_3$PbI$_3$ perovskites (MAPbI$_3$). Since the efficiency of c-Si cells is limited by Auger recombination due to particularly weak absorption in Si~\cite{richterReassessmentLimitingEfficiency2013}, there is very limited scope for efficiency increase by improvement of bulk material properties in c-Si (Fig.~\ref{fig:fom_real_materials}). If the effective $\alpha$ of silicon could be substantially increased (e.g. by extreme light trapping or polymorphism \cite{garnettLightTrappingSilicon2010,Fadaly2020}) without compromising on other properties, the $\Gamma_\mathrm{PV}$ FOM correctly predicts that
c-Si cells may achieve higher efficiencies by being less constrained by Auger recombination. 

Multicrystalline silicon (mc-Si), the CsPbI$_3$ inorganic perovskite, Cu(In,Ga)Se$_2$ chalcopyrites (CIGSe), and Cd(Se,Te) all have FOMs within an order of magnitude of each other (\SIrange{0.4}{2.6}{}) and have all achieved record efficiencies close to their FOM-predicted limit at their current stage of bulk material quality. On the other hand, CdTe absorbers without Se alloying, Sn-based hybrid perovskites (\ce{FASnI3}), and complex kesterite alloys based on the \ce{Cu2ZnSn(S,Se)4} composition have reached a much lower fraction of their current efficiency potential in spite of a similar FOM range to the previous group of absorbers. While the case of CdTe may simply be due to an overall community switch to alloyed Cd(Se,Te), the relatively high FOMs of kesterite and Sn-based perovskites likely indicate recent improvement in material quality in these systems~\cite{wangVacancyenhancedCationOrdering2026,shiInterfacialDipolesBoost2024} that has only partially translated to higher PV efficiencies. According to Fig.~\ref{fig:fom_real_materials}, kesterites and Sn-based perovskites have reached comparable bulk material quality to CIGSe chalcopyrites and \ce{CsPbI3} perovskites, respectively. Their current $\eta_\mathrm{\Gamma}$ limits are predicted to be 23.1\% (kesterites) and 26.7\% (\ce{FASnI3}).

Many PV absorbers fall in an intermediate \SIrange{0.005}{0.04}{} $\Gamma_\mathrm{PV}$ range (Fig.~\ref{fig:fom_real_materials}). In this FOM range, record efficiencies as a fraction of the $\eta_\mathrm{\Gamma,sq}$ limit seem to correlate with accumulated experience, quantified by number of PV-related publications for each material. Indeed, non-alloyed kesterites and hydrogenated amorphous silicon (a-Si) have reached higher efficiencies than \ce{Zn3P2}, Cu$_2$BaSn(S,Se) (CBTSSe) and BiOI; and no working solar cells have yet been reported for the most recently proposed materials in this FOM range (the \ce{BaZrS3} sulfide perovskite and the \ce{BaCd2P2} Zintl phosphide). The only exception to this rule is Se, which is the only material in our dataset to have reached both a relatively high value of $\Gamma_\mathrm{PV}$ and a high efficiency with respect to its $\eta_\mathrm{\Gamma,sq}$ limit, in spite of a low number of PV-relevant publications. One could speculate that this anomaly may be associated with the uniqueness of Se as a simple elemental, low-melting-point semiconductor, facilitating its processability~\cite{nielsenReemergenceSeleniumSolar2026}.

Among the absorbers that fall in the low FOM region ($\Gamma_\mathrm{PV} < 0.001$) two cases are particularly instructive. SnS has a long history of PV-related work, but its FOM is still below \SI{e-4}{}, indicating that its efficiency can't be improved much further without acting on its bulk properties, and that its prospects as a high-efficiency PV absorber are questionable. Second, the figure of merit of the recently proposed antiperovskite Ag$_3$SI is still so low ($ < \SI{e-7}{}$) that even a perfect device structure and perfect interfaces would not help this absorber reach meaningful efficiencies at this stage of bulk material quality.

Overall, the data in Fig.~\ref{fig:fom_real_materials} offer an intriguing insight: The experimental material quality of newly synthesized PV materials depends strongly on the material itself, but only weakly on accumulated experience on that specific material. Experience plays a bigger role later, in the processes of improving efficiencies and of increasing the FOM of an emerging material to the level of the best PV absorbers. In fact, the four materials in our dataset with 10 publications or less (\ce{Ag3SI}, \ce{Sn2SbS2I3}, \ce{NaBiS2} and \ce{BaCd2P2}) have figures of merit spanning all the way from \SI{e-8}{} (\ce{Ag3SI}) to \SI{2e-2}{} (\ce{BaCd2P2}). These findings suggest that, in the very early development stages of novel PV materials, a research strategy based on diversification and rapid experimental screening may be more effective than a strategy focusing on detailed investigation of fewer materials.

\begin{figure}[t!]
\centering%
\includegraphics[width=1.0\columnwidth]{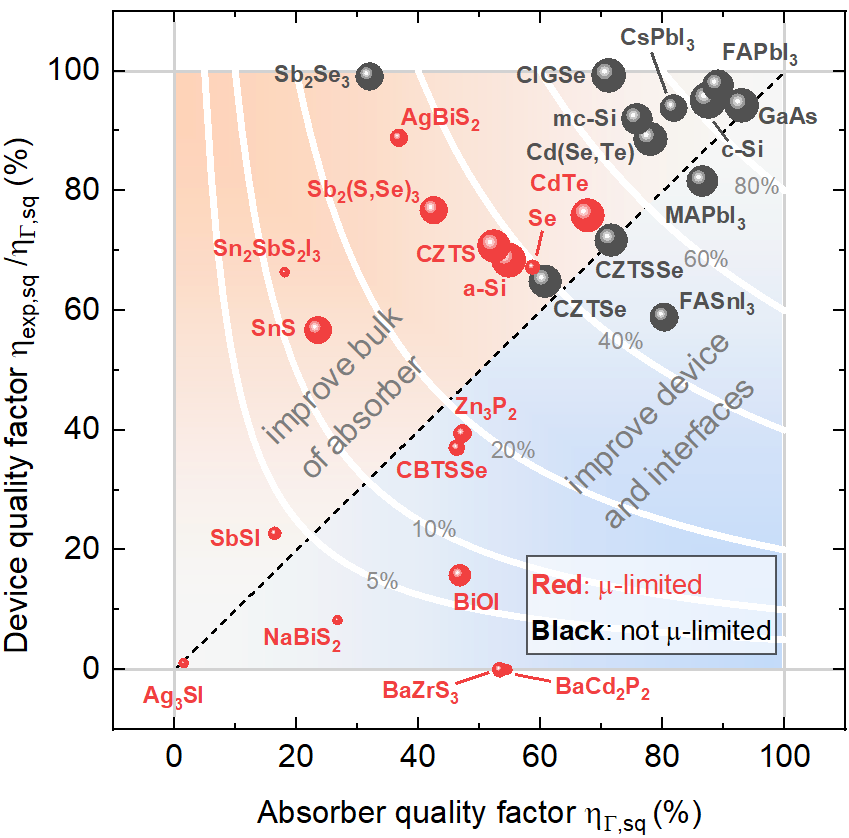}
\caption{Quantification of the current quality of 28 PV absorbers at the device/interface level (y-axis) and at the bulk absorber property level (x-axis). Along the dashed line the device quality factor is equal to the absorber quality factor. For materials located above this line, improving the bulk properties of the absorber is likely more urgent than improving the device structure, interfaces, contact/transport layers in their solar cells. For materials located below the dashed line, absorber development is likely less urgent. The white contour lines indicate device/absorber quality factor combinations with a constant $\eta_\mathrm{exp,sq}$ efficiency limit (labeled as fraction of the SQ limit). Red data points indicate materials that would have a $> 10\%$ (relative) higher efficiency limit if their carrier mobility could be increased from their current experimental value to infinity. The other materials (black points) would benefit less from a mobility increase.}
\label{fig:4_quadrants}
\end{figure}

The combination of experimental record efficiencies $\eta_\mathrm{exp,sq}$ and the $\Gamma_\mathrm{PV}$ figure of merit enables researchers to quantify the relative urgency of device-level optimization and bulk absorber-level optimization for a given single-junction PV technology. In Fig.~\ref{fig:4_quadrants}, a device quality factor defined as $\eta_\mathrm{exp,sq}/\eta_\mathrm{\Gamma,sq}$ is plotted against an absorber quality factor simply defined as $\eta_\mathrm{\Gamma,sq}$. For some PV technologies (e.g., FASnI$_3$, BiOI, NaBiS$_2$, BaZrS$_3$, BaCd$_2$P$_2$) the absorber quality factor is substantially higher than the device quality factor. In these cases, optimization of device structure, interfaces, contact and/or transport layers is arguably more urgent than improvement of the bulk properties of the absorber. For other technologies (e.g., CIGSe, Sb$_2$S$_3$, FAPbI$_3$, AgBiS$_2$) the device quality factor is close to 100\% and improving the absorber quality seems to be the only realistic way to significantly improve their current record efficiency.

By changing the $\mu$ values of the different PV materials (with everything else fixed) and applying Eqs.~\ref{eq:fom_definition},~\ref{eq:eta_from_fom}, it is also possible to determine which PV materials are limited by low carrier mobilities.
It turns out that over half of the materials shown in Figs.~\ref{fig:fom_real_materials},~\ref{fig:4_quadrants} would have an efficiency limit higher by at least 10\% relative if their carrier mobilities could be increased to infinity. These are examples of PV materials for which the classic detailed balance analysis would overestimate efficiency limits.

\begin{figure}[t!]
\centering%
\includegraphics[width=1.0\columnwidth]{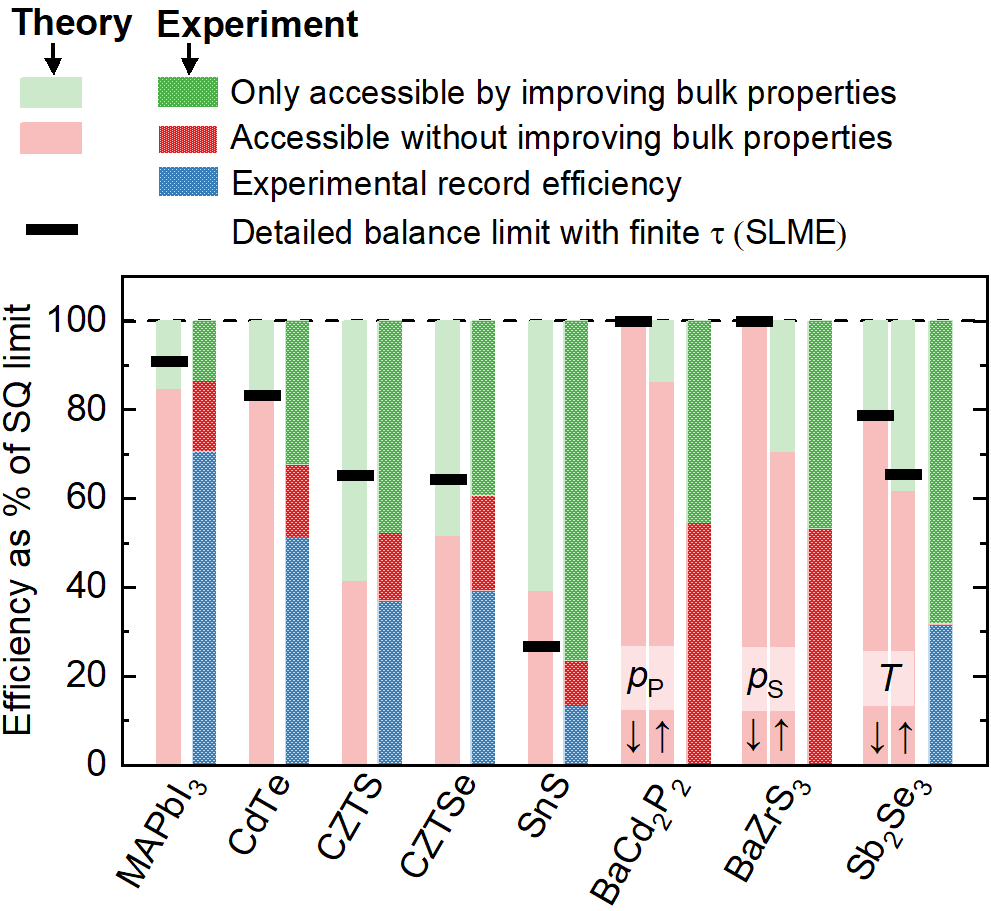}
\caption{Maximum efficiencies achievable by 8 PV absorbers, according to the $\Gamma_\mathrm{PV}$ FOM using properties calculated from first principles (pale colors) and experimentally measured properties on state-of-the-art specimens (bright colors). These efficiency limits are compared to classic detailed-balance limits ("SLME"~\cite{Yu2012,Blank2017}, see main text) including non-radiative recombination but excluding the effect of finite values of $\mu$, $n$, and $\epsilon$. The top of the red bars corresponds to the $\eta_\mathrm{\Gamma,sq}$ limit.
The top of the blue bars is the current record efficiency. The green bars represent the gap in efficiency between the $\eta_\mathrm{\Gamma,sq}$ limit and the SQ limit. $p_\mathrm{P}$ and $p_\mathrm{S}$ refer to the partial pressures of phosphorus and sulfur during growth (upward arrow: high pressure; downward arrow: low pressure). $T$ refers to the growth temperature of \ce{Sb2Se3} (upward arrow: \SI{375}{\celsius}; downward arrow: \SI{277}{\celsius}).
The material properties used to obtain $\Gamma_\mathrm{PV}$ from DFT modeling are listed in Table~S3, Supporting Information, including two additional materials not shown in the figure.}
\label{fig:bars}
\end{figure}

A key question that naturally arises at this point is whether it is actually possible to improve the bulk properties of a given absorber beyond its current state of the art. For example, is there any hope of developing SnS beyond its modest 23\% absorber quality factor? Could we have foreseen the significant recent FOM increase of kesterite absorbers? Can perovskite absorbers become even better? To search for answers, we can turn to first-principles calculations of material properties.
Until recently, only six out of the eight bulk properties necessary to calculate the FOM ($\alpha$, $\sigma$, $n$, $\epsilon$, $m$, $E_\mathrm{g}$) were within reasonable reach of materials modeling methods based on density functional theory (DFT). However, recent methodological advances~\cite{Ganose2021,Alkauskas2014} now allow computational scientists to also estimate $\tau$ by calculating carrier capture coefficients of point defects~\cite{Kim2020}, as well as $\mu$ by efficient calculation of carrier scattering rates~\cite{Dahliah2021}.

Hence, it is also possible to calculate the $\Gamma_\mathrm{PV}$ FOM and the corresponding $\eta_\mathrm{\Gamma,sq}$ efficiency limit using computationally-determined properties, even for the case of materials that have never been synthesized. Some examples are shown in Fig.~\ref{fig:bars}. The DFT-derived $\eta_\mathrm{\Gamma,sq}$ limits (pale colors) in materials with significant accumulated experience (MAPbI$_3$, CdTe, CZTS, CZTSe, SnS) are generally close to the corresponding limits derived from experimental properties (bright colors).
This is an important finding, because it implies that the ultimate efficiency potential of these absorbers could have been predicted (at least semi-quantitatively) before commencing any experimental work. Some key caveats on the comparison between the efficiency potential of a material based on experimental vs. calculated properties are given in Sec.~S5 of the SI.

Two other results in Fig.~\ref{fig:bars} are worth noting. First, the recently proposed BaZrS$_3$ and BaCd$_2$P$_2$ may be able to reach the SQ limit under anion-poor growth conditions according to the $\Gamma_\mathrm{PV}$ FOM based on calculated properties~\cite{yuanDiscoveryZintlphosphideBaCd2P22024,yuanAssessingCarrierMobility2024a}. BaCd$_2$P$_2$ is predicted to be compatible with high efficiencies even under P-rich conditions, emphasizing a particularly high level of defect tolerance in this material.
Second, efficiency limits calculated with the $\Gamma_\mathrm{PV}$ FOM can be compared to the corresponding limits calculated with current state-of-the-art detailed-balance methods including non-radiative recombination (finite $\tau$) but no collection losses (infinite $\mu$)~\cite{Yu2012,Blank2017,Kirchartz2018,kimInitioCalculationDetailed2021}. For computationally modeled absorbers, these methods are often referred to as the "spectroscopically limited maximum efficiency" (SLME) following the theoretical treatment in Ref.~\cite{Yu2012}. As expected, the $\eta_\mathrm{\Gamma,sq}$ efficiency limit is equal or more stringent than the SLME limit in all cases, with the exception of SnS. This discrepancy can be rationalized based on a significant difference in the absorption coefficient spectra used to calculate the SLME and the $\Gamma_\mathrm{PV}$ FOM for the specific case of SnS (see Sec.~S5).

\begin{figure}[t!]
\centering%
\includegraphics[width=1.0\columnwidth]{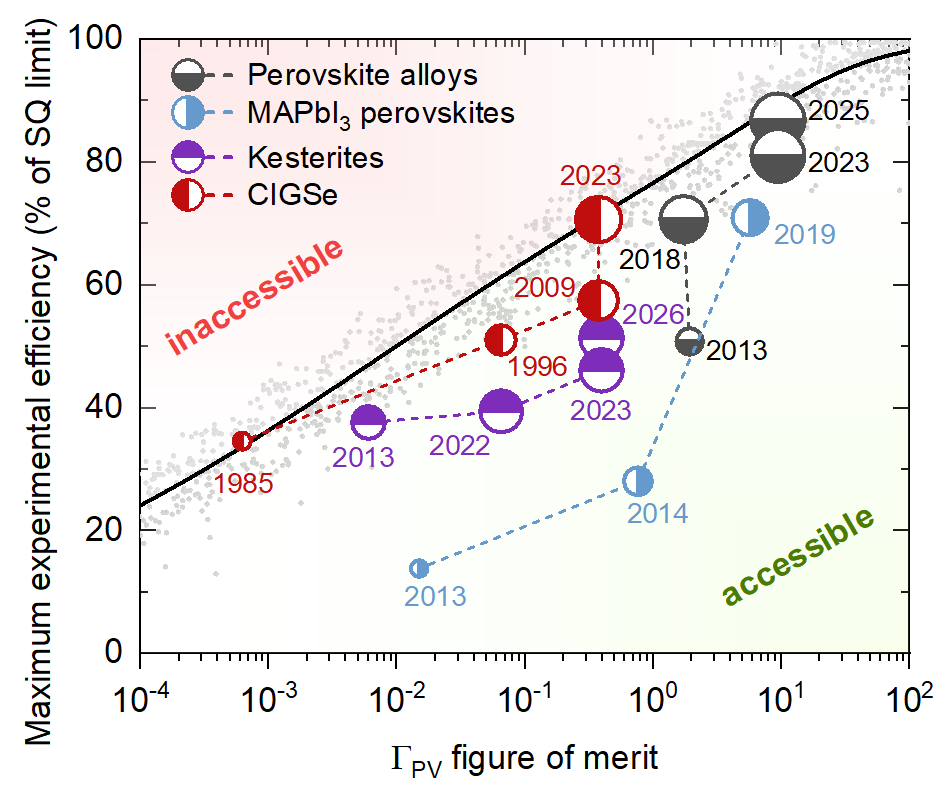}
\caption{Development history of four classes of PV absorbers. Kesterites include absorbers with the general (Ag,Cu)$_2$(Cd,Zn)Sn(S,Se)$_4$ formula; CIGSe stands for Cu(In,Ga)Se$_2$; MAPbI$_3$ perovskites only include materials without alloying. Perovskite alloys are absorbers with the general (MA,FA,Cs)Pb(I,Br,Cl)$_3$ formula. Numbers next to each data point refer to the year of publication. The marker size is proportional to the logarithm of the number of cumulated publications for that material class in the year indicated next to the marker. The thick black line is the $\eta_\mathrm{\Gamma,sq}$ efficiency limit (Eq.~\ref{eq:eta_from_fom}, same as in Fig.~\ref{fig:fom_real_materials}). The small grey data points are the training set used to derive the $\Gamma_\mathrm{PV}$ FOM and giving an impression of the error bar of the $\eta_\mathrm{\Gamma,sq}$ limit. Raw data is listed in Tables~S5--S7, Supporting Information.}
\label{fig:history}
\end{figure}

Finally, the $\Gamma_\mathrm{PV}$ FOM can be used to track the progress of different PV absorbers and their record solar cells through history. These timelines are plotted in Fig.~\ref{fig:history} for four PV absorber types and can be related to technological breakthroughs in absorber processing and to device structure/interface innovations. As an example, let us follow the history of CIGSe (data shown in Table~S6). Record CIGSe cells between 1985 and 1996 were operating near the maximum efficiency allowed by the absorber's bulk quality at that time. In this period, the FOM of CIGSe improved due to increases in $\tau$ and $n$ following the introduction of a Cu-rich growth phase, as well as Na and Ga incorporation~\cite{Scheer2011}. Steady improvements to the three-stage evaporation process in CIGSe over the next decade~\cite{Scheer2011} resulted in significantly longer carrier lifetimes and a large increase in the $\Gamma_\mathrm{PV}$ FOM. However, the device quality as of 2009 was no longer sufficient for CIGSe cells to achieve their full potential. The main breakthrough allowing CIGSe solar cells to again approach their $\eta_\mathrm{\Gamma,sq}$ efficiency limit in the 2020's was the introduction of an alkali post-deposition treatment (2013) to improve the charge-separating heterointerface~\cite{Chirila2013}.

\section{Conclusion}

The $\Gamma_\mathrm{PV}$ FOM can be considered a generalization of previously proposed $\alpha \tau$ and $\alpha \tau \mu$ FOMs for PV materials. It is applicable across a wide range of realistic combinations of bulk properties of possible PV photoabsorber materials (Fig.~\ref{fig:different_foms}). The efficiency limit $\eta_\mathrm{\Gamma}$ set by the inductive $\Gamma_\mathrm{PV}$ FOM is more stringent than the corresponding limit set by deductive methods based on the principle of detailed balance. The main reason is that efficiency losses by imperfect carrier collection (primarily influencing $J_\mathrm{sc}$ and FF) are taken into account in $\Gamma_\mathrm{PV}$ by including the carrier mobility, doping density, and dielectric constant of the PV absorber as relevant properties. Applying the $\Gamma_\mathrm{PV}$ FOM to 28 established and emerging PV materials (Fig.~\ref{fig:fom_real_materials}) allows us to quantitatively visualize their current development status, to decide whether to prioritize bulk absorber improvement or device-level improvement (Fig.~\ref{fig:4_quadrants}), and to track the impact of different innovations in absorber processing versus device/interface processing and design (Fig.~\ref{fig:history}). With some caveats, the $\eta_\mathrm{\Gamma}$ efficiency limit obtained using first-principles-calculated properties seems to be an acceptable descriptor of the ultimate performance level that may be expected from candidate PV absorbers (Fig.~\ref{fig:bars}). This gives credibility to computational screening workflows that include calculation of the eight properties necessary to evaluate the $\Gamma_\mathrm{PV}$ FOM. Intriguingly, certain exotic compounds recently identified via such methods exhibit higher efficiency potential than many established PV absorbers.

\section*{Acknowledgements}
This work was funded by the European Union (ERC, IDOL, 101040153). Views and opinions expressed are however those of the author only and do not necessarily reflect those of the European Union or the European Research Council. Neither the European Union nor the granting authority can be held responsible for them. This work was supported by a research grant (42140) from VILLUM FONDEN.

\vspace{3cm}

\onecolumngrid
\section*{Data availability}
Information supporting the results presented in this paper can be found in a separate preprint available on \texttt{https://chemrxiv.org/doi/abs/10.26434/chemrxiv.15006235/v1}. This separate document includes extended sections on: Criteria and priorities applied to choose values of the material properties used to evaluate $\Gamma_\mathrm{PV}$; Common pitfalls to be avoided when choosing such properties; Notes on comparing the efficiency potential of a material from experimental and calculated properties; Specific remarks on each of the 304 properties used for the 38 materials presented in this work.
The results presented in this paper can be reproduced with either a Jupyter notebook or an Excel spreadsheet, both of which are available on \texttt{https://github.com/DTU-Nanolab-materials-discovery/PV-figure-of-merit}. The $\Gamma_\mathrm{PV}$ FOM framework has been integrated in the \textsc{solphin} open-source package available on \texttt{https://github.com/SMTG-Bham/solphin}.


\bibliography{MyLibrary3}

\end{document}